\documentclass[useAMS,usenatbib]{mn2e}
\usepackage{txfonts,graphicx,natbib}
\bibpunct{(}{)}{;}{a}{}{,}
\begin{document}
\title[Spectral and light curve evolution of SN 2008ax]{The type IIb SN 2008ax: spectral and light curve evolution}
\author[Pastorello et al.]{A. Pastorello$^{1}$\thanks{E-mail:a.pastorello@qub.ac.uk}, M. M. Kasliwal$^{2,3}$, 
R. M. Crockett$^{1}$, S. Valenti$^{1}$, R. Arbour$^{4}$, \and K. Itagaki$^{5}$, S. Kaspi$^{6,7}$, A. Gal-Yam$^{8}$, S. J. Smartt$^{1}$, 
R. Griffith$^{9,10}$, K. Maguire$^{1}$, \and
E. O. Ofek$^{2}$, N. Seymour$^{11}$, D. Stern$^{10}$, W. Wiethoff$^{12}$ \\
$^{1}$ Astrophysics Research Centre, School of Mathematics and Physics, Queen's University Belfast, Belfast BT7 1NN, UK\\
$^{2}$ Division of Physics, Mathematics, and Astronomy, 105-24, California Institute of Technology, Pasadena, CA 91125, US\\
$^{3}$ Hale Fellow, Gordon and Betty Moore Foundation \\
$^{4}$ Pennell Observatory, 29 Wrights Way, South Wonston, Hants, SO21 3HE, UK \\
$^{5}$ Itagaki Astronomical Observatory, Teppo-cho, Yamagata 990-2492, Japan\\
$^{6}$ School of Physics and Astronomy and the Wise Observatory, Raymond and Beverly Sackler Faculty of Exact Sciences, Tel Aviv University, Tel Aviv 69978, Israel\\
$^{7}$ Physics Department, Technion, Haifa 32000, Israel\\
$^{8}$ Benoziyo Center for Astrophysics, Weizmann Institute of Science, 76100 Rehovot, Israel\\
$^{9}$ Department of Astronomy, University of California, Berkeley, CA 94720-3411, US\\
$^{10}$ Jet Propulsion Laboratory, California Institute of Technology, Pasadena, CA 91109, US\\
$^{11}$ Spitzer Science Center, California Institute of Technology, 1200 East California Boulevard, Pasadena, CA 91125, US\\
$^{12}$ Port Wing, Wisconsin 54865, US\\
}
\date{Accepted .....; Received ....; in original form ....}

\maketitle
\label{firstpage}

\begin{abstract}
We present spectroscopy and photometry of the He-rich supernova (SN) 2008ax.
The early-time spectra show prominent P-Cygni H lines, which decrease with time and 
disappear completely about two months after the explosion. In the same period He I lines become the 
most prominent spectral features. SN 2008ax displays the ordinary spectral evolution of a type IIb 
supernova. A stringent pre-discovery limit
constrains the time of the shock breakout of SN 2008ax to within only a few hours. 
Its light curve, which peaks in the B band about 20 days after the explosion, 
strongly resembles that of other He-rich core-collapse supernovae. The observed evolution of
SN 2008ax is consistent with the explosion of a young Wolf-Rayet (of WNL type) star, which had retained a thin, 
low-mass shell of its original H envelope. The overall characteristics of SN 2008ax 
are reminiscent of those of SN 1993J, except for a likely smaller H mass. This may account 
for the findings that the progenitor of SN 2008ax was a WNL star and not a K supergiant as in the
case of SN 1993J, that a prominent early-time peak is missing in the light curve of SN 2008ax,
and that H$\alpha$ is observed at higher velocities in SN 2008ax than in SN 1993J.   

\end{abstract}


\section{Introduction}  \label{intro}

In recent years, the attention of the astronomical community toward H-poor 
core-collapse (CC) supernovae (SNe) has significantly increased.  This is due 
to the discovery, during the late 90s, of the connection between 
H- and He-stripped CC SNe (of type Ic) and long duration gamma-ray bursts \citep[GRBs, see e.g.][]{gal98,woo06}.
However, interest in He-rich events (type Ib) has been rather marginal so far, 
although recently invigorated by the discovery of early X-ray
emission from the type Ib SN 2008D \citep[SN 2008D,][]{sol08,mal08}.
 %mazz08 
 Equally poorly studied are the very few 
SN events in which there is spectroscopic evidence for the presence of H lines in 
otherwise normal type Ib SN spectra. These rare CC SNe were dubbed as type IIb by \citet{woo87} \citep[see also][]{fil97},
and their hybrid spectroscopic appearance was interpreted as the result of the explosion of 
a He core of an originally massive star, which had retained a residual (though marginal) 
H envelope (few $\times$ 10$^{-1}$M$_\odot$) at the time of explosion. 

The explosion of the very nearby SN 1993J offered the unique opportunity to study in 
detail the evolution of a type IIb event and to detect the progenitor star in pre-explosion 
images \citep{ald94,coh95,van02}, although it would be more correct to 
state that the binary system producing SN 1993J was observed in the pre-SN images, 
since signatures of the presence of a massive, blue companion star arose from spectro-photometric 
observations of the SN site obtained about ten years after the SN explosion \citep{mau04}.
These exceptional data constrained the precursor of SN 1993J to be a K supergiant 
of initial mass around 14M$_\odot$, but with a He core of only 3-6M$_\odot$ at the time of the explosion
\citep[see also][]{nom93,pod93,woo94}. 

Another type IIb event, SN 2001ig, showed periodic modulation
in its radio light curve. This may indicate either density enhancements indicative of circumstellar shells produced
in the thermal-pulsing phase by a single asymptotic giant branch star or, more likely, a stellar wind modulated by
motion in an eccentric binary system\footnote{\protect\citet{sor06} also explained the observed modulation 
in the radio light curves of the type IIb SNe 2001ig and 2003bg in terms of density enhancements due to quasi-periodic mass loss episodes 
from massive WR progenitors, occurred soon prior to the SN explosions.} \citep{ryd04}. The binary system scenario
is also supported by spectropolarimetric observations of SN 2001ig \citep{mau07}.
Interestingly, a point-like source was detected
at the SN position about 1000 days after the explosion, possibly the massive (10-18M$_\odot$) companion of 
the Wolf-Rayet (WR) star producing SN 2001ig \citep{ryd06}.

Apart from SN 1993J \citep[e.g.][]{ric94,lew94,bar95}, extensive data sets have been published 
for only a very limited number of He-rich CC SNe,
including SNe 1987K \citep{fil88}, 1990I \citep{elm04}, 1996cb \citep{qiu99},
1999dn \citep{deng00,ben08}, 1999ex \citep{str02,ham02}, 2008D \citep{sol08}, 
plus the peculiar SNe 1991D \citep{ben02}, 2001gh \citep{nancy08} and 2005bf \citep{anu05,tom05,fol06}.
SN 2006jc and similar events \citep[the so-called SNe Ibn, see][]{pasto08a} are different because
the He is mostly confined in the circumstellar environment \citep{mat00,pasto07,fol07,smi08,imm08,mat08,tom08,pasto08b}.

SN 2008ax is therefore the second He-rich CC SN that has excellent
spectro-photometric monitoring, and for which we can derive directly
information on the progenitor, through the analysis of deep pre-explosion
archive images \citep{cro08}. In this paper we present spectroscopy and photometry
of SN 2008ax, while in the companion paper \citep{cro08} we study
in detail the nature of the SN precursor.
 
\begin{figure}
\centering
\includegraphics[width=8.4cm]{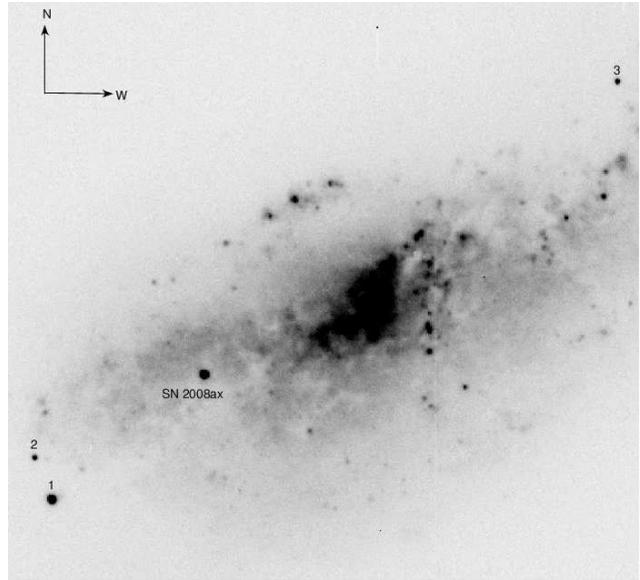}
\caption{SN 2008ax in NGC 4490. V band image obtained on March 22, 2008, with the
Liverpool Telescope (equipped with RatCAM). Three photometric comparison stars 
are also labelled. \label{fig1}}
\end{figure}

\section{SN 2008ax and its host galaxy}

SN 2008ax (see Fig. \ref{fig1}) was discovered independently by \citet[][on March 3.45 UT]{mos08} and 
\citet[][on March 4.62 UT]{nak08} in the nearby barred spiral galaxy (type SBcd) 
NGC 4490. The new object was around 16th magnitude at discovery. The coordinates were 
$\alpha$ = $12^{h}30^{m}40\fs80$ and $\delta$ = $+41\degr38\arcmin14\farcs5$,
quite far (53$\farcs$1 East and 25$\farcs$8 South) from the center of the host galaxy
\citep{mos08}. Remarkably, the explosion site was monitored by \citet{arb08} about 
6 hours (JD=2454528.7) before the detection of \citet{mos08} (JD 2454528.95) and 
the images show no sign of the SN. This allows us to constrain the time of the shock breakout to high
precision, with an uncertainty of only a few hours. Hereafter in the paper we will adopt
JD = 2454528.80 $\pm$ 0.15 as the time of the shock breakout, and all phases will be
computed with reference to this epoch.

NGC 4490 and NGC 4485 are a well known pair of late-type, interacting galaxies 
embedded in an extended and asymmetric H I envelope \citep{huc80}, which is elongated perpendicularly
to the plane of NGC 4490. \citet{cle98} claim that the configuration of the 
neutral hydrogen envelope might result from a bipolar outflow of H I driven by 
starburst activity / SN explosions, and not from the interaction
between the two companion galaxies. This is in agreement with the particularly high star formation rate 
estimated in  NGC 4490 \citep{via80,thr89,cle99}.

The distance modulus ($\mu$) for NGC 4490 is not well constrained. From \cite{tully88} and assuming a Hubble constant 
H$_0$ = 72 km s$^{-1}$ Mpc$^{-1}$, we would obtain $\mu$ = 29.55 magnitudes. However, \citet{rob02} noted that
\cite{tully88} gave significantly inconsistent distance estimates  for the two interacting galaxies,
9.7 Mpc for NGC 4485 and 8.1 Mpc for NGC 4490 (with our choice of H$_0$). This inconsistency
in Tully's estimate for the two companion galaxies suggests a need to compute the distance to NGC 4490
also using different approaches. 
An attempt to estimate the distance of the SN host galaxy was performed by \citet{ter02}. It was determined from
the technique of sosie galaxies \citep{pat94} making use of a number of different calibrators for which the
distances were computed from two independent Cepheid calibrations. This method gives $\mu$ = 29.90 $\pm$ 1.16 
magnitudes (d = 9.5 Mpc). A further possibility is to derive the distance via the recessional velocity of the SN host galaxy. 
LEDA\footnote{http://leda.univ-lyon1.fr/} \citep{pat03} provides a recessional velocity for NGC 4490 corrected 
for Local Group infall into the Virgo Cluster of v$_{Vir}$ = 797 km s$^{-1}$. From this, we obtain a
distance d of $\sim$11.1 Mpc, corresponding to a distance modulus $\mu$ = 30.22 magnitudes. 
If we consider the recessional velocity with respect to the CMB background (v$_{3k}$ = 817 km s$^{-1}$), 
we obtain $\mu$= 30.27 magnitudes (d = 11.3 Mpc). A lower distance is estimated
from the recessional velocity corrected to the centroid of the Local Group, v$_{lg}$ = 618 km s$^{-1}$,
which gives $\mu$= 29.67 magnitudes (d = 8.6 Mpc).
 Averaging these 5 estimates to minimize the uncertainty, we obtain 
$\mu$= 29.92 $\pm$ 0.29 magnitudes (i.e. d = 9.6 Mpc) which will be used throughout this paper.\footnote{A consistency 
check can be done by considering the pair NGC 4485 and NGC 4490 as members of the 14-4 group. 
An accurate distance of the group was obtained by \citet{tu08} 
through a weighted average of 4 different methods tied to the HST Cepheid Key Project scale \citep{fri01}. Rescaling 
for consistency to H$_0$ = 72 km s$^{-1}$ Mpc$^{-1}$, a distance modulus 
$\mu$= 29.91 $\pm$ 0.13 was obtained. This distance for the 14-4 group (d $\approx$ 9.6 Mpc) is fully consistent with our average estimate
for NGC 4490.}

\begin{figure}
\centering
\includegraphics[width=8.9cm]{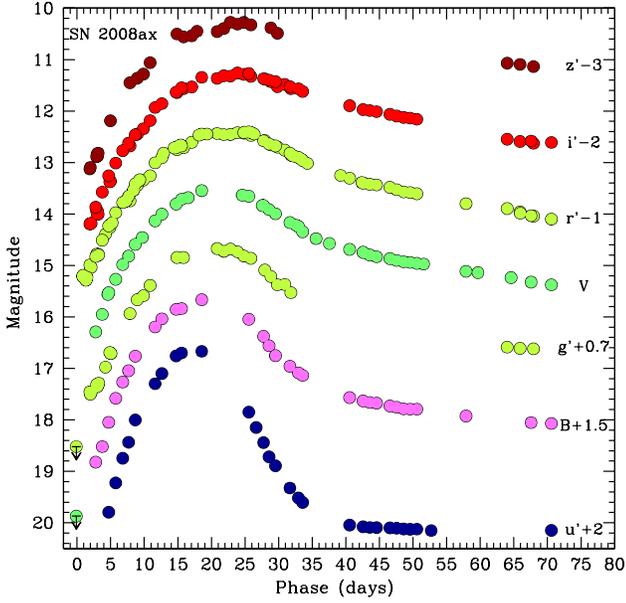}
\caption{u'BVg'r'i'z' light curves of SN 2008ax until 2 months after the explosion. The phase is computed since
shock breakout. \label{fig2}}
\end{figure}

The reddening due to the Galaxy in the direction of NGC 4490
is very small \citep[E(B-V)=0.022 magnitudes,][]{sch98}. However, there is evidence of additional
reddening inside the host galaxy (see discussion in Sect. \ref{sect_spec}) and therefore
we will adopt the value E(B-V) = 0.3 magnitudes as the best estimate for the total extinction toward SN 2008ax.

The faint absolute magnitude measured at discovery (around -13.4) led \citet{mos08} to erroneously suggest
that the SN was possibly an highly extinguished event or, alternatively, the super-outburst of a 
luminous blue variable. Instead, it was the first light coming from the explosion of 
one of the youngest stripped-envelope CC SNe ever discovered.

\begin{table*}
\footnotesize
\caption{Optical photometry of SN 2008ax. \label{tab1}}
\begin{tabular}{cccccccccc}
\hline\hline
Date & JD & u' & B & V  & g' & r' & i' & z' & Instrument \\
\hline
Mar 05 & 2454530.70 &               &              &              & 16.756 0.053 & 15.990 0.039 & 16.194 0.055 & 16.127 0.133 & P60 \\ 
Mar 05 & 2454530.79 &               &              &              & 16.802 0.089 & 16.034 0.076 & 16.185 0.038 & 16.093 0.031 & P60 \\ 
Mar 05 & 2454530.90 &               &              &              & 16.765 0.076 & 16.037 0.055 & 16.184 0.117 &              & P60 \\ 
Mar 06 & 2454531.58 &               & 17.323 0.210 & 16.290 0.067 & & 15.826 0.056 & 15.867 0.055 & &  LT\\
Mar 06 & 2454531.70 &               &              &              & 16.615 0.030 & 15.792 0.050 & 15.946 0.085 & 15.884 0.056 & P60 \\ 
Mar 06 & 2454531.79 &               &              &              & 16.657 0.139 & 15.781 0.048 & 15.997 0.126 & 15.837 0.087 & P60 \\ 
Mar 06 & 2454531.87 &               &              &              & 16.634 0.098 & 15.806 0.103 & 15.950 0.065 & 15.874 0.118 & P60 \\ 
Mar 06 & 2454531.96 &               &              &              & 16.614 0.033 & 15.785 0.039 & 16.006 0.105 & 15.817 0.026 & P60 \\ 
Mar 06 & 2454532.04 &               &              &              & 16.587 0.060 &   &   &   & P60 \\ 
Mar 07 & 2454532.55 &               & 17.021 0.032 & 15.952 0.042 & & 15.509 0.024 & 15.576 0.027 & &  LT\\
Mar 07 & 2454533.03 &               &              &              & 16.277 0.102 & 15.395 0.129 &   &   & P60 \\ 
Mar 08 & 2454533.53 &  17.794 0.069 & 16.548 0.029 & 15.528 0.014 & & 15.233 0.019 & 15.258 0.017 & &  LT\\
Mar 08 & 2454533.69 &               &              &              & 15.992 0.072 & 15.265 0.109 & 15.377 0.099 &   & P60 \\ 
Mar 08 & 2454533.78 &               &              &              & 16.004 0.053 & 15.206 0.085 & 15.359 0.104 & 15.190 0.085 & P60 \\ 
Mar 08 & 2454533.86 &               &              &              & 16.012 0.079 &   &   &   & P60 \\ 
Mar 09 & 2454534.57 &  17.226 0.040 & 16.082 0.017 & 15.270 0.011 & & 14.974 0.012 & 15.012 0.011 & &  LT\\
Mar 10 & 2454535.60 &  16.746 0.035 & 15.766 0.018 & 14.983 0.012 & & 14.778 0.012 & 14.771 0.009 & &  LT\\
Mar 10 & 2454536.49 &  16.435 0.036 & 15.545 0.021 & 14.819 0.013 & & 14.654 0.016 & 14.624 0.014 & &  LT\\
Mar 11 & 2454536.68 &               &              &              & 15.236 0.043 & 14.745 0.227 & 14.678 0.087 & 14.451 0.080 & P60 \\ 
Mar 11 & 2454537.47 &  16.002 0.034 & 15.264 0.014 & 14.592 0.011 & & 14.416 0.013 & 14.461 0.013 & &  LT\\
Mar 12 & 2454537.77 &               &              &              & 14.965 0.084 & 14.443 0.025 & 14.451 0.095 & 14.362 0.131 & P60 \\ 
Mar 13 & 2454538.68 &               &              &              & 14.888 0.059 & 14.342 0.034 & 14.344 0.067 & 14.283 0.124 & P60 \\ 
Mar 14 & 2454539.68 &               &              &              & 14.689 0.053 & 14.257 0.034 & 14.186 0.075 & 14.061 0.085 & P60 \\ 
Mar 14 & 2454540.42 &  15.298 0.028 & 14.695 0.013 & 14.133 0.010 & & 14.007 0.012 & 13.932 0.010 & &  LT\\
Mar 15 & 2454541.43 &  15.099 0.031 & 14.538 0.015 & 14.005 0.010 & & 13.907 0.014 & 13.856 0.013 & &  LT\\
Mar 18 & 2454543.55 &  14.759 0.031 & 14.354 0.016 & 13.808 0.010 & & 13.710 0.012 & 13.633 0.011 & &  LT\\
Mar 18 & 2454543.67 &               &              &              & 14.145 0.181 & 13.749 0.102 & 13.641 0.132 &              & P60 \\ 
Mar 18 & 2454544.36 &  14.700 0.040 & 14.338 0.014 & 13.714 0.010 & & 13.677 0.011 & 13.546 0.010 & &  LT\\
Mar 19 & 2454544.66 &               &              &              & 14.147 0.050 & 13.714 0.083 & 13.564 0.062 & 13.561 0.052 & P60 \\ 
Mar 20 & 2454545.87 &               &              &              &              & 13.608 0.061 & 13.533 0.037 & 13.537 0.068 & P60 \\ 
Mar 21 & 2454546.66 &               &              &              &              &              &              & 13.449 0.056 & P60 \\ 
Mar 21 & 2454547.36 &  14.669 0.300 & 14.165 0.020 & 13.546 0.012 & & 13.448 0.011 & 13.345 0.010 & &  LT\\
Mar 24 & 2454549.65 &               &              &              & 13.971 0.050 & 13.436 0.030 & 13.366 0.069 & 13.456 0.338 & P60 \\ 
Mar 25 & 2454550.65 &               &              &              & 14.025 0.122 & 13.458 0.094 & 13.315 0.126 & 13.396 0.228 & P60 \\ 
Mar 26 & 2454551.64 &               &              &              & 13.981 0.019 & 13.454 0.149 & 13.322 0.075 & 13.274 0.046 & P60 \\ 
Mar 27 & 2454552.68 &               &              &              & 14.034 0.038 & 13.420 0.033 & 13.259 0.123 & 13.312 0.238 & P60 \\ 
Mar 28 & 2454553.64 &               &              &              & 14.117 0.049 & 13.429 0.092 & 13.287 0.144 & 13.276 0.095 & P60 \\ 
Mar 28 & 2454554.37 &  15.849 0.029 & 14.545 0.015 & 13.657 0.009 & & 13.406 0.011 & 13.270 0.010 & &  LT\\
Mar 29 & 2454554.66 &               &              &              & 14.161 0.062 & 13.471 0.047 & 13.328 0.098 & 13.325 0.087 & P60 \\ 
Mar 29 & 2454555.47 &  16.149 0.026 &              &              & &              &              & &  LT\\
Mar 31 & 2454556.56 &  16.441 0.025 & 14.877 0.012 & 13.837 0.009 & & 13.578 0.011 & 13.371 0.009 & &  LT\\
Mar 31 & 2454556.75 &               &              &              & 14.391 0.111 & 13.574 0.056 & 13.380 0.110 &              & P60 \\ 
Mar 31 & 2454557.37 &  16.719 0.034 & 15.060 0.014 & 13.915 0.010 & & 13.642 0.011 & 13.398 0.015 & &  LT\\
Apr 01 & 2454557.64 &               &              &              & 14.515 0.049 & 13.658 0.118 & 13.432 0.127 & 13.383 0.041 & P60 \\ 
Apr 01 & 2454558.34 &  16.892 0.064 & 15.254 0.021 & 13.996 0.013 & & 13.680 0.018 & 13.428 0.018 & &  LT\\   
Apr 02 & 2454558.63 &               &              &              & 14.676 0.097 & 13.701 0.020 & 13.528 0.082 & 13.489 0.115 & P60 \\ 
Apr 03 & 2454559.70 &               &              &              & 14.669 0.049 & 13.759 0.122 & 13.482 0.035 &              & P60 \\ 
Apr 03 & 2454560.48 &  17.325 0.034 & 15.460 0.014 & 14.168 0.009 & & 13.821 0.013 & 13.525 0.018 & &  LT\\
Apr 04 & 2454560.63 &               &              &              & 14.828 0.074 & 13.856 0.109 & 13.566 0.068 &              & P60 \\ 
Apr 05 & 2454561.75 &  17.518 0.032 & 15.587 0.013 & 14.244 0.010 & & 13.902 0.012 & 13.567 0.011 & &  LT\\
Apr 05 & 2454562.38 &  17.606 0.069 & 15.636 0.017 & 14.346 0.012 & & 13.962 0.012 & 13.621 0.011 & &  LT\\
Apr 12 & 2454569.37 &  18.044 0.056 & 16.066 0.014 & 14.687 0.012 & & 14.305 0.011 & 13.895 0.009 & &  LT\\  
Apr 14 & 2454571.37 &  18.078 0.044 & 16.133 0.017 & 14.741 0.011 & & 14.401 0.012 & 13.972 0.010 & &  LT\\ 
Apr 15 & 2454572.37 &  18.090 0.089 & 16.162 0.018 & 14.796 0.012 & & 14.423 0.014 & 13.992 0.039 & &  LT\\
Apr 16 & 2454573.38 &  18.095 0.032 & 16.177 0.014 & 14.832 0.011 & & 14.436 0.013 & 14.012 0.015 & &  LT\\
Apr 18 & 2454575.36 &  18.099 0.062 & 16.231 0.021 & 14.867 0.010 & & 14.482 0.011 & 14.065 0.010 & &  LT\\
Apr 19 & 2454576.37 &  18.106 0.041 & 16.255 0.015 & 14.903 0.011 & & 14.528 0.013 & 14.096 0.011 & &  LT\\
Apr 20 & 2454577.40 &  18.120 0.043 & 16.288 0.016 & 14.930 0.010 & & 14.571 0.011 & 14.124 0.010 & &  LT\\
Apr 21 & 2454578.37 &  18.126 0.043 & 16.293 0.019 & 14.951 0.011 & & 14.586 0.016 & 14.147 0.013 & &  LT\\ 
Apr 22 & 2454579.38 &  18.124 0.038 & 16.293 0.017 & 14.961 0.011 & & 14.607 0.011 & 14.160 0.016 & &  LT\\
Apr 25 & 2454581.50 &  18.151 0.070 &              &              & &              &              & &  LT\\ 
Apr 30 & 2454586.68 &               & 16.425 0.038 & 15.114 0.015 & & 14.800 0.014 &              & &  LT\\ 
May 06 & 2454592.82 &               &              &              & 15.888 0.037 & 14.896 0.038 & 14.548 0.049 & 14.070 0.056 & P60 \\ 
May 08 & 2454594.72 &               &              &              & 15.910 0.052 & 14.958 0.045 & 14.590 0.056 & 14.094 0.058 & P60 \\ 
May 09 & 2454596.37 &               & 16.554 0.020 & 15.324 0.010 & & 15.033 0.012 & 14.583 0.010 & &  LT\\ 
May 10 & 2454596.73 &               &              &              & 15.920 0.034 & 15.041 0.102 & 14.627 0.053 & 14.137 0.041 & P60 \\ 
May 12 & 2454599.38 & 18.149 0.031  & 16.571 0.017 & 15.376 0.010 & & 15.100 0.011 & 14.611 0.012 & &  LT\\
\hline
\end{tabular}

LT = 2-m Liverpool Telescope + RatCAM; P60 = Robotic Palomar 60-inch Telescope + CCD
\end{table*}

\section{The light curve} \label{sect_photo}

\begin{table}
\footnotesize
\caption{Unfiltered photometry of SN 2008ax from amateur astronomers, 
rescaled to the Johnson-Bessell V and/or Sloan r' bands. 
\label{tab2}}
\begin{tabular}{ccccc}
\hline\hline
Date & JD & CV  & Cr' & Observer \\ \hline
Mar 03  & 2454528.70 & $>$19.87     & $>$19.52     & RA \\
Mar 04  & 2454529.67 &              & 16.210 0.188 & WW \\
Mar 04  & 2454529.68 &              & 16.220 0.200 & WW \\
Mar 04  & 2454529.69 &              & 16.192 0.195 & WW \\
Mar 04  & 2454530.12 &              & 16.272 0.090 & KI \\
Mar 04  & 2454530.12 &              & 16.285 0.127 & KI \\
Mar 04  & 2454530.13 &              & 16.282 0.109 & KI \\
Mar 04  & 2454530.13 &              & 16.277 0.127 & KI \\
Mar 04  & 2454530.14 &              & 16.271 0.078 & KI \\
Mar 04  & 2454530.15 &              & 16.264 0.110 & KI \\
Mar 04  & 2454530.15 &              & 16.259 0.088 & KI \\
Mar 04  & 2454530.15 &              & 16.261 0.145 & KI \\
Mar 04  & 2454530.16 &              & 16.255 0.079 & KI \\
Mar 04  & 2454530.18 &              & 16.252 0.108 & KI \\
Mar 04  & 2454530.25 &              & 16.249 0.104 & KI \\
Mar 04  & 2454530.25 &              & 16.239 0.084 & KI \\
Mar 04  & 2454530.29 &              & 16.229 0.141 & KI \\
Mar 06  & 2454532.03 &              & 15.795 0.167 & KI \\
Mar 07  & 2454533.41 & 15.562 0.031 &              & RA \\
Mar 08  & 2454534.04 &              & 15.172 0.135 & KI \\
Mar 10  & 2454536.05 &              & 14.751 0.083 & KI \\
Mar 10  & 2454536.06 &              & 14.747 0.109 & KI \\
Mar 11  & 2454537.21 &              & 14.536 0.114 & KI \\
Mar 11  & 2454537.21 &              & 14.535 0.099 & KI \\
Mar 12  & 2454538.01 &              & 14.356 0.136 & KI \\
Mar 12  & 2454538.13 &              & 14.334 0.129 & KI \\
Mar 13  & 2454538.51 & 14.457 0.040 &              & RA \\
Mar 16  & 2454541.67 &              & 13.818 0.121 & WW \\
Mar 17  & 2454543.24 &              & 13.718 0.126 & KI \\
Mar 18  & 2454544.07 &              & 13.669 0.112 & KI \\
Mar 18  & 2454544.25 &              & 13.660 0.129 & KI \\
Mar 19  & 2454545.35 & 13.683 0.118 &              & RA \\
Mar 21  & 2454546.97 &              & 13.468 0.113 & KI \\
Mar 22  & 2454548.15 &              & 13.448 0.096 & KI \\
Mar 27  & 2454553.39 & 13.634 0.043 &              & RA \\
Mar 28  & 2454553.60 &              & 13.420 0.127 & WW \\
Mar 28  & 2454553.72 &              & 13.413 0.100 & WW \\
Mar 29  & 2454554.69 &              & 13.415 0.123 & WW \\
Mar 29  & 2454555.17 &              & 13.448 0.118 & KI \\
Mar 30  & 2454556.36 & 13.836 0.055 &              & RA \\
Apr 04  & 2454560.84 &              & 13.863 0.086 & WW \\
Apr 04  & 2454561.09 &              & 13.877 0.075 & KI \\
Apr 04  & 2454561.40 & 14.219 0.060 &              & RA \\
Apr 05  & 2454562.16 &              & 13.935 0.106 & KI \\
Apr 06  & 2454563.07 &              & 14.017 0.063 & KI \\
Apr 07  & 2454564.38 & 14.479 0.030 &              & RA \\
Apr 09  & 2454566.35 & 14.575 0.044 &              & RA \\
Apr 11  & 2454568.02 &              & 14.251 0.074 & KI \\
Apr 14  & 2454571.18 &              & 14.387 0.121 & KI \\
Apr 14  & 2454571.41 & 14.750 0.046 &              & RA \\ 
Apr 23  & 2454580.39 & 14.973 0.074 &              & RA \\ 
May 01  & 2454588.47 & 15.140 0.048 &              & RA \\ 
May 06  & 2454593.40 & 15.237 0.026 &              & RA \\ 
May 08  & 2454594.78 &              & 14.985 0.059 & WW \\
\hline
\end{tabular}

RA = 40cm f/5 Newtonian + SXV-H9 CCD (R. Arbour, UK)\\
WW = 35cm C14 OTA + DSI Pro II w/Sony EXview HAD CCD (ICX 429) (W. Wiethoff, US)\\
KI = 60cm f/5.7 + Bitran BT-214E CCD (Kodak KAF-1001E) (K. Itagaki, Japan)\\

%% Any table notes must follow the \end{tabular} command.
\end{table}

\begin{figure*}
\centering
\includegraphics[width=15.3cm]{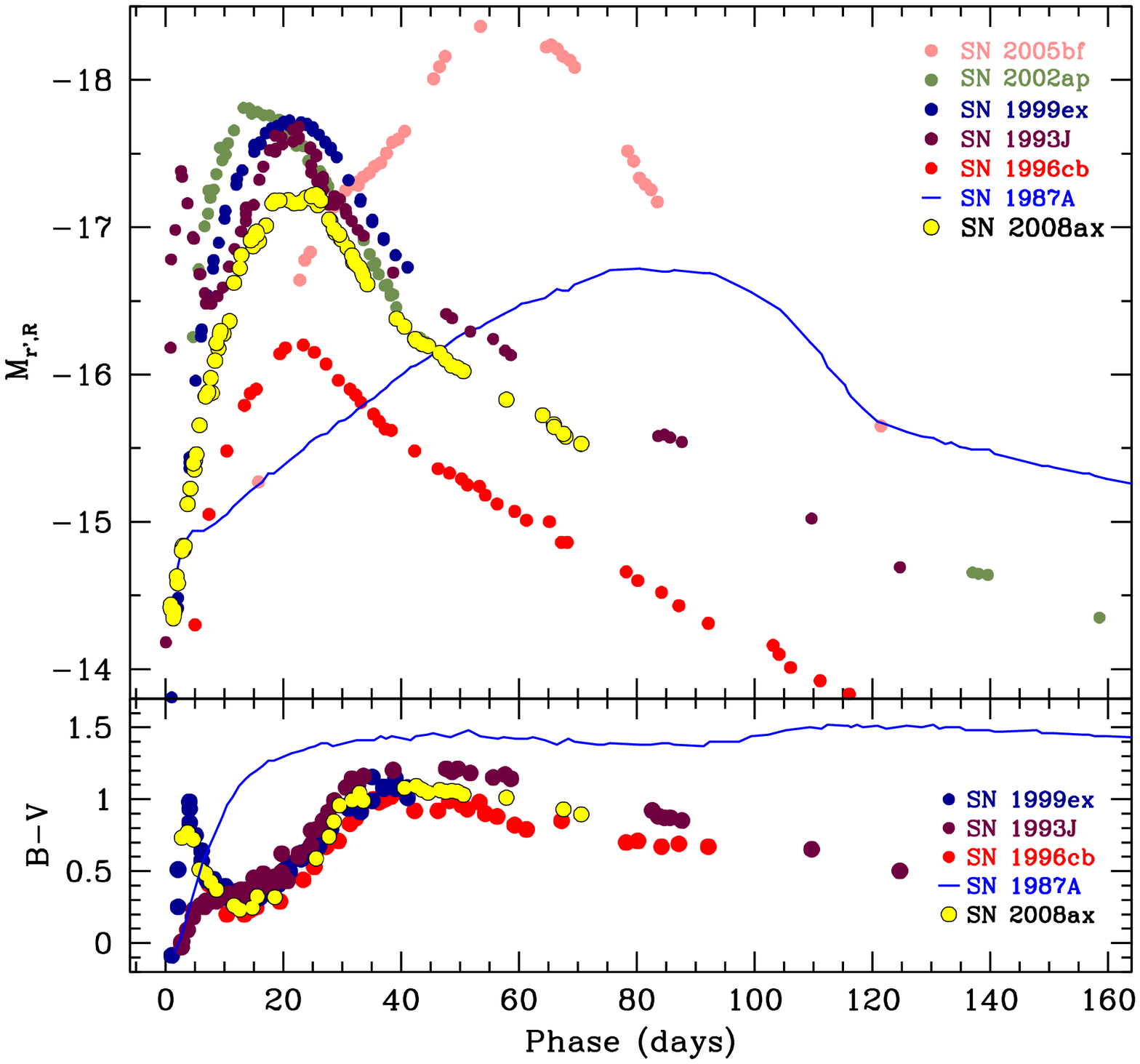}
\caption{Top: r'-band absolute light curve of SN 2008ax, compared with the red band light
curves of the peculiar type Ib SN 2005bf \protect\citep{fol06}, the broad-line type Ic SN 2002ap
\protect\citep{gal02,yoz03,fol03,tom06}, the normal type Ib SN~1999ex \protect\citep{str02}, the type IIb SNe 1993J \protect\citep{lew94,bar95} and 
1996cb \protect\citep{qiu99}, and the peculiar type IIP SN 1987A 
\protect\citep{menzies87,catchpole87,catchpole88}.
 Bottom: B-V colour evolution for SNe 2008ax, 1987A, 1996cb, 1993J, 1999ex.
\label{fig3}}
\end{figure*}

SN 2008ax is one of the best ever monitored core-collapse SNe, starting
with a deep pre-explosion image obtained only 6 hours prior to the
SN discovery of \citet{mos08}. 
The observational campaign of SN 2008ax started soon after the discovery and covered
a period of about 70 days. Photometry has been obtained at the 2-m Liverpool Telescope \citep{ste04}
in La Palma  (Canary Islands, Spain) 
and the 60-inch Telescope of the Palomar Observatory \citep{cen06}. In addition, unfiltered data 
collected by  amateur astronomers have been used in our analysis. These data have been
scaled to the Johnson-Bessell V-band or the Sloan r'-band photometry, depending on the 
sensitivity curves of the CCDs used in these observations. We find that the quantum efficiency (QE)  
curve of the SXV-H9 CCD used by R.A. peaks around 5100\AA, so the unfiltered magnitudes of this images
are best compared to the V-band magnitudes, while both the DSI Pro II  and the KAF-1001E cameras 
used by W.W. and K.I., respectively, have CCDs with QEs peaking around 6000-6200\AA. 
Therefore, the unfiltered magnitudes obtained with these two imagers are scaled to
the Sloan r' magnitudes.

\begin{table}
\centering
\caption{Epoch of maximum, rise time to maximum and peak magnitude
for the u'BVr'i' light curves of SN 2008ax
\label{tab3}}
\begin{tabular}{cccc}
\hline\hline
Band & JD(max) & Rise time (days) & M($\lambda$)$_{max}$  \\ \hline
u' & 2454546.2$\pm$0.4 & 17.4 & 14.65$\pm$0.02 \\
B  & 2454547.7$\pm$0.5 & 18.9 & 14.16$\pm$0.03 \\
g' & 2454549.3$\pm$0.6 & 20.5 & 13.95$\pm$0.02 \\
V  & 2454549.5$\pm$0.5 & 20.7 & 13.51$\pm$0.02 \\
r' & 2454551.1$\pm$0.5 & 22.3 & 13.38$\pm$0.03 \\
i' & 2454551.6$\pm$0.6 & 22.8 & 13.26$\pm$0.02 \\
z' & 2454553.1$\pm$1.0 & 24.3 & 13.28$\pm$0.03 \\
\hline
\end{tabular}
\end{table}
 
\begin{table*}
\footnotesize
\caption{Log of spectroscopic observations.
\label{tab4}}
\begin{tabular}{ccccccc}
\hline\hline
Date & JD & Phase  & Instrumental configuration & Exposure time (s) & Spectral range (\AA) & Res. (\AA) \\ \hline
Mar 12 & 2454538.38 & 9.6 & Wise 40-inch + FOSC + g600 & 2$\times$3600 & 3990--8230 & 18 \\ 
Mar 18 & 2454544.35 & 15.6 & Wise 40-inch + FOSC + g600 & 2$\times$1800 & 3990--8240 & 18 \\
Mar 29 & 2454554.64 & 25.8 &  Palomar 200-inch + TSPEC & 5$\times$200 & 9500--13500,14050--18150,19230--24670 & $\ddag$ \\
Mar 31 & 2454556.70 & 27.9 & Palomar 200-inch + DBSP +g300+g316 & 2$\times$60+2$\times$120 & 3100--5480,5650--8170 & 9.5,13 \\
Apr 01 & 2454557.69 & 28.9 & Palomar 200-inch + DBSP +g300+g316 & 3$\times$60             & 3100--5470,5650--8170 &  6,7 \\
Apr 02 & 2454558.70 & 29.9 & Palomar 200-inch + DBSP +g300+g316 & 3$\times$60             & 3100--5460,5650--8170 & 7.5,9 \\
Apr 08 & 2454565.34 & 36.5 & Wise 40-inch + FOSC + g600 & 3600         & 4010--8210 & 18 \\
Apr 28 & 2454584.64 & 55.8 & Palomar 200-inch + DBSP +g600+g316 & 2$\times$200             & 3130--5590,5640--9690& 3.5,9.5 \\
\hline
\end{tabular}

$\ddag$ 1.2 \AA~in the J band, 2.9 \AA~in the K band
\end{table*}

Johnson-Bessell B and V, and Sloan u', g', r', i' and z' photometry of SN 2008ax is presented
in Tab. \ref{tab1}, while the unfiltered amateurs' data are in Tab. \ref{tab2}. The resulting light
curves (including the calibrated amateurs data) are shown in Fig. \ref{fig2}. 
Like in many H-stripped core-collapse SNe the peak luminosity in the blue bands is reached 
a few days earlier than in the red (about 19 days after the shock breakout in the B band and 23 in the i' band, see Tab. \ref{tab3}). 
After maximum, the light curves
decline rapidly in all bands until day $\sim$40, when the SN luminosity settles 
onto the radioactive tail.

H-poor core-collapse supernovae display a wide range of behaviour in their light curve evolution.
In Fig. \ref{fig3} (top) we compare the r'-band absolute light curve of SN 2008ax with those (Sloan r' or 
Johnson-Bessell R) of a number of H-poor SNe (see caption of Fig. \ref{fig3} for details). 
In particular, we note that the light curve of 
SN 2008ax is remarkably different from that of the peculiar type II SN 1987A (that shares
some early-time spectroscopic similarity with SN 2008ax, see Sect. \ref{sect_spec}). It is
instead similar to those of the H-stripped core-collapse events shown in Figure \ref{fig3}. The overall shape
resembles that of the type IIb SNe 1996cb and 1993J. In particular, its peak magnitude 
(M$_R$ $\approx$  -17.3 magnitudes) is quite similar to that of SN 1993J (the difference is $\Delta$M$_R \approx$ 0.3), but
is significantly brighter (by $\sim$ 1 magnitude) than that of SN~1996cb. A major difference between the two SNe is that SN 2008ax does not show the
prominent and narrow early-time peak exhibited by the light curve of SN 1993J. This peak is 
attributed to the initial shock heating and the subsequent cooling of a 
low-mass envelope \citep[see e.g.][]{bar94,shi94}. In SN 2008ax only a marginally 
detectable shoulder during the first day after shock breakout was observed, in analogy to that reported by \citet{str02}
in the type Ib SN 1999ex. This possibly due to SN 2008ax having an initial radius that was smaller 
than that of SN 1993J.

In Fig. \ref{fig3} (bottom) we also compare the colour evolution of SN 2008ax with those of the 
type II SN 1987A, the type Ib SN 1999ex and the type IIb SNe 1993J and 1996cb. SN 1987A has a different 
colour evolution compared to the other three objects, that evolve in a rather similar fashion 
(apart from the early-time blue excess due to the shock breakout visible in SN 1993J). 
In particular, during the first 10 days SN 2008ax becomes rapidly bluer, moving 
from B-V = 0.8 to 0.2 (note that the colour evolution of SN 1993J follows the opposite trend in this phase).
During the subsequent period, SN 2008ax (and the other type Ib/IIb events in Fig. \ref{fig3}) becomes 
redder, reaching B-V = 1.1 at phase $\sim$ 40 days. Then, the colour 
becomes slowly bluer again with time.

\section{The metamorphosis of  SN 2008ax} \label{sect_spec}

\begin{figure}
\centering
\includegraphics[width=9.25cm]{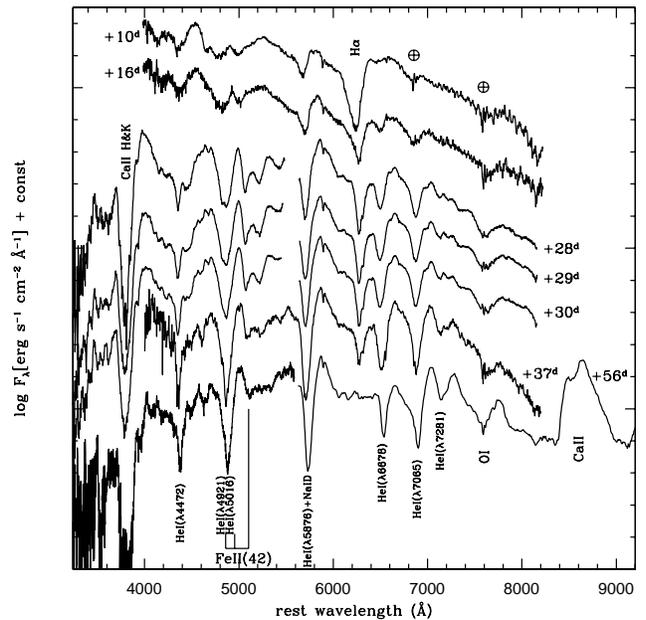}
\caption{Spectral evolution of SN 2008ax. The most significant spectral 
lines and the epochs since core-collapse of the various spectra are also labelled. 
The spectra, corrected for total reddening E(B-V)=0.3, are in the host galaxy rest frame.
The positions of the most important telluric absorptions are also marked, with $\oplus$.
\label{fig4}}
\end{figure}

SN 2008ax showed an amazing spectral evolution, transforming in a few weeks
through many different spectral types. \citet{blo08} initially classified it as
a young 1987A-like type II SN. The strong interstellar Na I D absorption
visible in the spectrum of \citet{blo08} also suggested significant host galaxy 
reddening, E(B-V) = 0.6. The presence of prominent H and He I lines, with
blueshifted peaks, was indicative of very rapid spectral evolution, another
characteristic in common with SN 1987A.
The most important difference was the much broader spectral lines in SN 2008ax, 
corresponding to ejecta velocities in the range 23000-26000
km s$^{-1}$. In SN 1987A they were lower by a factor 2/3. 

However, subsequent spectra showed increasing Fe II, Ca II and, most significantly, He I
features, suggesting that this SN should be re-classified as a type IIb \citep{cho08}. 
Also, based on a new measurement of the EW of the Na I interstellar doublet ($\sim$0.18nm), 
\citet{cho08} used an unspecified method to revise the estimate of the host galaxy contribution
to the reddening to E(B-V) = 0.5 magnitudes.

\citet{mar08}, analysed a near-infrared (NIR) spectrum of SN 2008ax, and noted the
presence of prominent He I features at 1.08 and 2.06 microns and 
a weak Paschen $\beta$. These features make this spectrum similar to the early-time NIR spectra of
SN Ib 1999ex \citep{ham02}. The transition of SN 2008ax toward a type Ib event 
was also revealed by optical spectra collected by \citet{tau08}. 
However, \citet{tau08} pointed out that some H was still visible in their spectra,
although less prominent than the stronger He I lines.

\begin{figure}
\centering
\includegraphics[width=9cm]{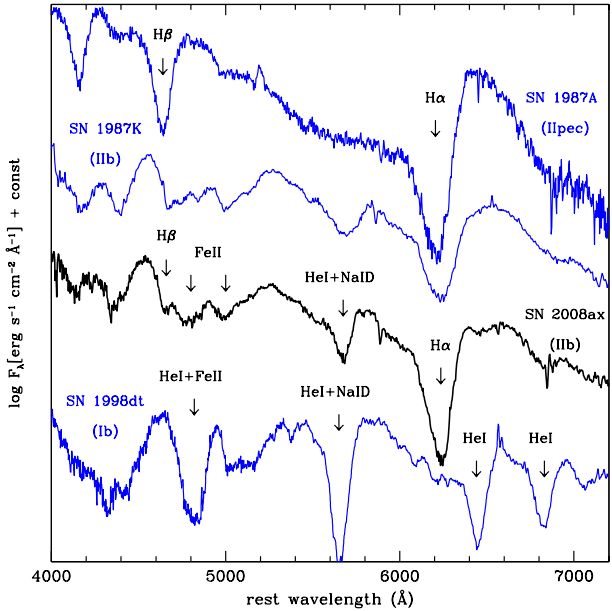}
\caption{Comparison between the early-time (March 12) spectrum  of SN 2008ax and those of the peculiar type IIP
SN 1987A \protect\citep{pun95}, the type IIb 
SN 1987K  \protect\citep{fil88} and the normal type Ib SN 1998dt \protect\citep{mat01}.
The spectrum of SN 1987K, which provides the best match to the earliest spectrum of SN 2008ax,
has been selected making use of the Superfit SN spectral identification code 
\protect\citep{how05}.
\label{fig5}}
\end{figure}

We collected a sequence of optical spectra of SN 2008ax  
using the 40-inch Telescope at the Wise Observatory (Israel)  
equipped with a Faint Object Spectrographic Camera (FOSC) and 
the 200-inch Hale Telescope at the Palomar Observatory (California, US) 
equipped with the Double Spectrograph \citep[DBSP,][]{oke82}. The log of spectroscopic observations is in Tab. \ref{tab4}.
All these spectra show relatively
strong interstellar Na ID, whose equivalent width (EW) is about 1.8\AA, in excellent agreement
with the value reported by \citet{cho08}. However, adopting the relation between EW and E(B-V) 
derived by \citet{tura03}, we obtain E(B-V)$\approx$ 0.3, which is significantly smaller than the
reddening estimate of \citet{blo08} and \citet{cho08}.

\begin{figure}
\centering
\includegraphics[width=9cm]{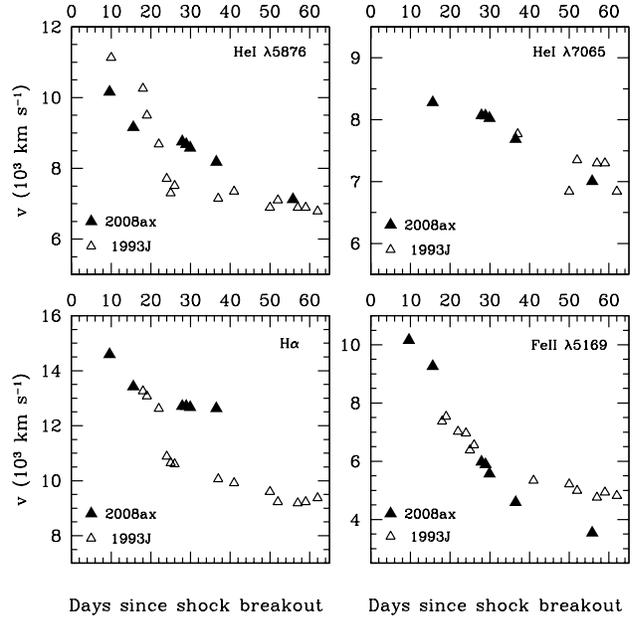}
\caption{Evolution of line velocities for a few selected lines in SN 2008ax and SN 1993J.
The data of SN 1993J are from  \protect\citet{bar95}.
\label{fig6}}
\end{figure}

\begin{figure}
\centering
\includegraphics[width=8.9cm]{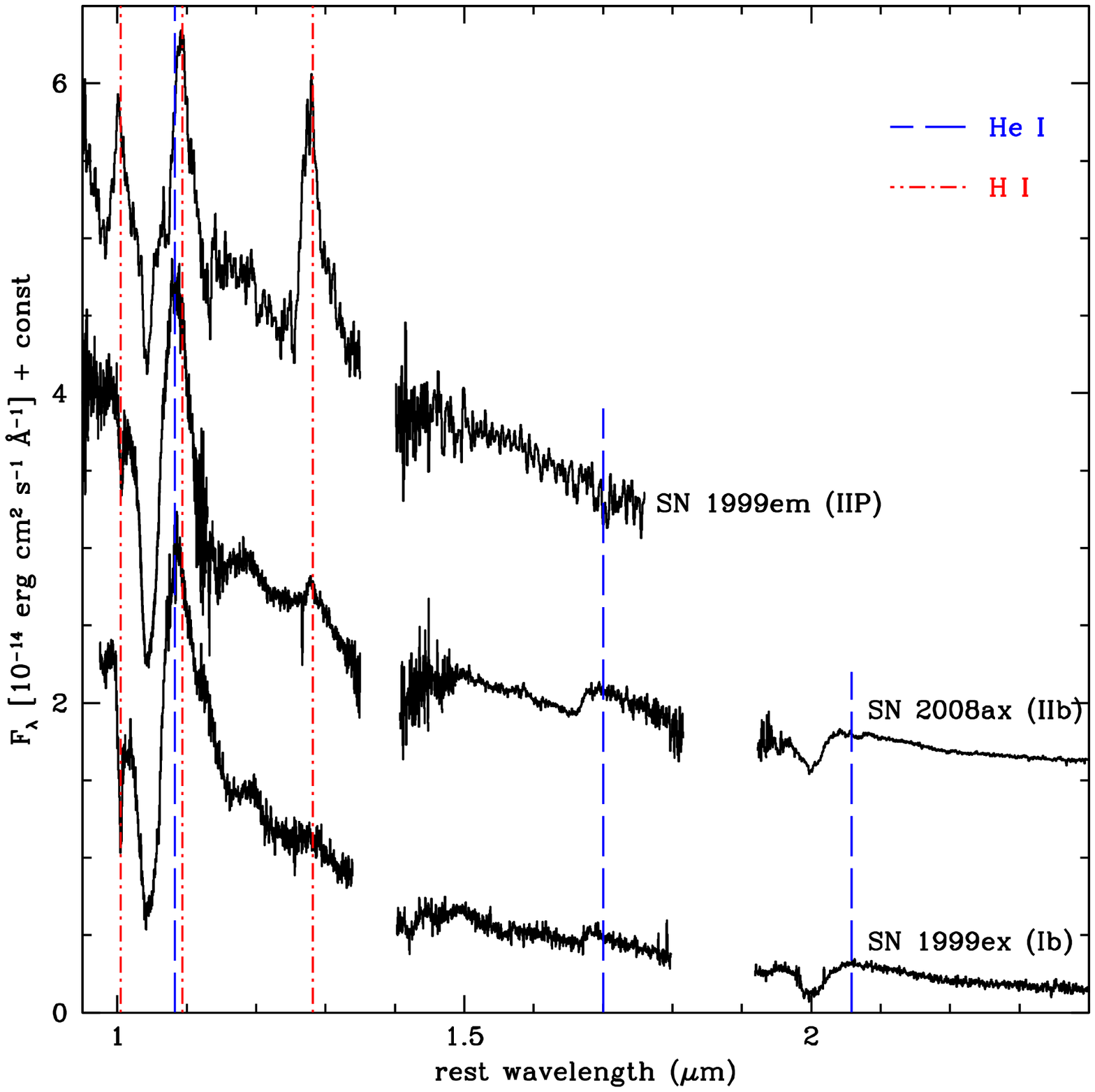}
\caption{Comparison between the NIR spectrum  of SN 2008ax obtained on Apr 29, and those of the type IIP
SN 1999em \protect\citep{elm03} and the normal type Ib SN 1999ex \protect\citep{str02}. The spectra of
SNe 2008ax and 1999ex have been reddening corrected, while that of SN 1999em has been artificially 
reddened to make the comparison clearer. All spectra are in the host galaxy rest frame. 
The vertical dashed blue lines mark the expected rest position of the main He I lines, while the dot-dashed red lines
mark the position of the Paschen $\delta$,  Paschen $\gamma$ and Paschen $\beta$ H lines.
\label{fig7}}
\end{figure}

The metamorphosis of SN 2008ax, whose spectrum transitioned through different 
spectral types (II$\rightarrow$IIb$\rightarrow$Ib) is  shown in Fig. \ref{fig4}, while a 
comparison with early spectra of the peculiar type II-plateau SN 1987A, the normal SN Ib 1998dt
and the type IIb event 1987K is shown in Fig. \ref{fig5} (two of these spectra have been downloaded from the SUSPECT
\footnote{http://bruford.nhn.ou.edu/$^{\sim}$suspect/index1.html} database). It is remarkable that the H$\alpha$ feature, 
which is strong in the early-time spectra of SN~2008ax, progressively weakens and disappears $\sim$2 months 
after core-collapse (Fig. \ref{fig4}). In this phase, the spectrum of SN 2008ax is that of a type Ib SN. 
SN 1987K experienced an analogous ''identity crisis'' \citep{fil88},
with the early-time spectra being rather similar to those of an H-rich type II SN
(e.g. 1987A, see Fig. \ref{fig5}) and very different from those of a canonical type Ib.
Some months later, however, SN 1987K evolved toward a spectrum dominated by Ca and O
forbidden lines and not containing the H$\alpha$ line in emission, which should be present in a type II SN in the nebular
phase. This wide-ranging spectral evolution seems to be quite common in type IIb SNe:
to a lesser extent, an analogous transition was seen in the spectra of SNe 1993J \citep{lew94,bar95}, 1996cb \citep{qiu99}
and 2001ig \citep{mau07}. 

The evolution of the line velocities, measured from the position of the P-Cygni minima of
He I 5876\AA, He I 7065\AA, H$\alpha$ and Fe II 5169\AA~is shown in Fig. \ref{fig6},
 both for SN 2008ax and SN 1993J \citep[data from][]{bar95}. It is remarkable that
 the velocities of the two type IIb SNe are comparable, although the velocity of 
 H$\alpha$, after the initial drop, remains much higher in SN 2008ax than in SN 1993J, while that of Fe II
 5169\AA~is lower in SN~2008ax than in SN~1993J. This, and the evidence that H$\alpha$
 is missing in the spectrum obtained two months after the explosion, suggest that the H envelope is
 probably less massive in SN~2008ax than in SN 1993J

SN 2008ax was also observed in the near-infrared (NIR) a few days past maximum (on March 29, 2008) with the Triple Spectrograph
\citep[TSPEC,][]{wil04}
mounted at the 200-inch Hale Telescope. This spectrum is compared with those of the type Ib SN 1999ex and
the type II-plateau SN 1999em about 3-4 weeks after the explosion in Fig. \ref{fig7}.
The most important H and He I lines in the NIR region are marked (the He I lines with 
dashed blue lines, the H lines with dot-dashed red lines).
The most prominent feature in all the spectra is a P-Cygni line peaking around 10900 \AA. In SN 1999em
this feature is likely due to a blend of Paschen $\gamma$ and He I \citep[with the possible contribution of C I,
see][]{pasto03}, while in SNe 2008ax and 1999ex it is mostly He I 10830\AA. Other He I lines identified in the spectra of
both SNe 1999ex and 2008ax are at about 17000\AA~and 20580\AA. In agreement with what we
find in the optical spectra, some H is also still visible in the NIR spectrum of SN 2008ax.
Paschen $\beta$, which was prominent in SN 1999em, is indeed visible (though weak) in SN 2008ax at $\lambda \approx$ 
12820\AA, while it was not definitely detected in the type Ib SN 1999ex. 
This is consistent with the classification of SN~2008ax as a type IIb event.

\section{Discussion}
Although the spectra of SN 2008ax are dominated by He I lines, the presence
of H features in the early-time spectra is unequivocal (see Fig. \ref{fig4} and Fig. \ref{fig7}), suggesting that
this SN should be classified as a type IIb event. We showed in Sect. \ref{sect_photo} and Sect. \ref{sect_spec} 
that SN 2008ax shares many similarities with other well studied SNe IIb (e.g. 1993J and 1996cb).
Nevertheless, the light curves of these SNe are also similar to those of totally 
H-stripped SNe Ib (see Fig. \ref{fig3}), indicating that the presence of a residual H skin does 
not significantly affect the luminosity evolution of SNe IIb. The most significant difference between the light curves of
type Ib/IIb SNe lies in their intrinsic luminosity, which is dependent on the amount of $^{56}$Ni ejected
in the explosion.

\begin{figure}
\centering
\includegraphics[width=8.9cm]{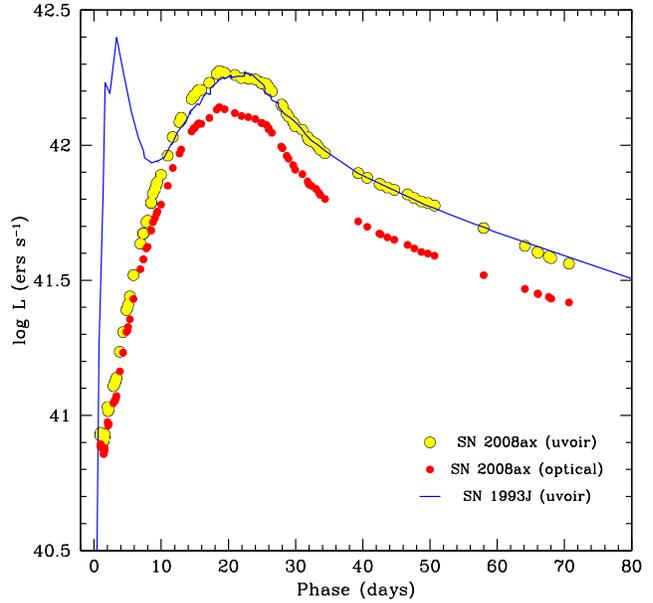}
\caption{Comparison between the quasi-bolometric ({\sl uvoir}) light curve of SN 2008ax (filled yellow symbols) with that of SN 1993J (solid blue line). 
Missing NIR (JHK) observations, the {\sl uvoir} light curve of SN 2008ax has been obtained integrating the fluxes in optical bands (from u' to z', red points)
and rescaled accounting the same NIR contribution as estimated for SN 1993J. 
\label{fig8}}
\end{figure}

In order to estimate the $^{56}$Ni mass synthesized by
SN~2008ax, its quasi-bolometric ({\sl uvoir}) light curve was computed and compared with that of SN 1993J.
The {\sl uvoir} light curve of SN 1993J was obtained by integrating the flux in the optical and NIR bands 
\citep[using the data of][]{lew94,bar95,wad97,matt02}. The observed optical light curve of SN 2008ax (Fig. \ref{fig8}, open red points)  
was rescaled, assuming for this object the same fractional NIR contribution to the {\sl uvoir} light curve as SN 1993J (solid blue line). The {\sl uvoir} light curve of SN 2008ax 
is shown in Fig. \ref{fig8} with filled yellow symbols. Surprisingly, there is an excellent agreement in the {\sl uvoir} luminosity 
between the two SNe. This is contrary to what we observed in individual bands (see e.g. Fig. \ref{fig3}), where SN~2008ax was slightly more luminous than SN 1993J 
in the blue bands and fainter in the red bands. The similar {\sl uvoir} luminosity  implies that the two objects have roughly
the same mass of ejected $^{56}$Ni. In SN~1993J the $^{56}$Ni mass was around 0.07-0.11$_\odot$ (see discussion below), and this
is also the $^{56}$Ni mass range expected for SN 2008ax.

There is also a similarity in the overall shape of the light curves of SN 2008ax and SN 1993J, 
with the exception of the initial strong contribution of the shock breakout to the early time light 
curve of SN 1993J. This raises the issue of whether the progenitor and the explosion parameters are similar in these two SNe. 
The width of the light curve during the photospheric phase and
the slope of the radioactive tail are known to depend on both the kinetic 
energy (E$_k$) and the ejected mass (E$_{ej}$) \citep{arn82}. In particular, the comparable widths of the light curve
peaks of SNe 1993J and 2008ax suggest a similar value for the M$_{ej}^3$/E$_k$ ratio \citep{arn82},
which, however, does not necessarily imply that M$_{ej}$ and E$_k$  are themselves similar.

Additional information on the individual values of M$_{ej}$ and E$_k$ could be obtained from the spectroscopy. 
The square of the photospheric velocity in envelope-stripped CC SNe is indeed proportional to the E$_k$/M$_{ej}$ ratio \citep{arn82}. 
Since the velocity of the Fe II 5169\AA~line is only marginally lower in SN 2008ax than in SN 1993J at the same phase 
(see Fig. \ref{fig6}, right-bottom panel), this may be an indication of comparable ejected 
masses and energies  in the two SNe.
Most theoretical papers on SN 1993J \citep[e.g.][]{nom93,pod93,woo94,bar94,utr94,shi94,you95} 
converge toward a standard-energy explosion (E$_k$ $\approx$ 10$^{51}$ erg) of a 3-6M$_\odot$ He core
with a residual H skin (few $\times$ 10$^{-1}$ M$_\odot$). From an inspection of the observed SN properties,
we therefore expect similar ejecta and explosion parameters for SN~2008ax.

An attempt to derive the ejecta mass and explosion energy  was performed for the type Ib SN 1999ex by
\citet{str02} using a comparison with the 6C hydrogenless model of \citet{woo87} for SNe Ib. Values
of  M$_{ej} \sim$  5-6M$_\odot$ and E$_k \approx$ 3 $\times$ 10$^{51}$ erg were derived for that SN. Recently,
\citet{sol08} obtained similar values for the type Ib SN 2008D: M$_{ej} \sim$  5M$_\odot$ and 
E$_k \approx$ 2 $\times$ 10$^{51}$ erg. However, the light curves of these two SNe are slightly broader
than that of 2008ax, whose shape, and hence likely its explosion and ejecta parameters, are probably 
closer to those of the type IIb SNe 1993J and 1996cb.

\citet{ald94}, from an analysis of deep pre-explosion images, found that the progenitor of the type IIb SN 1993J
was a K supergiant. The SN precursor was an originally massive (12-17M$_\odot$) member of a binary system,
in which the two companions had comparable main sequence masses \citep{mau04}.
In the case of SN 2008ax, the final configuration of the progenitor was likely a WR (WNL) star \citep[either a single, massive WR or 
a lower-mass WR in an interacting binary system, see][]{cro08} with
a residual H shell ($<$ 0.1M$_\odot$). The spectroscopic and photometric evolution of SN 2008ax agrees with this progenitor type.

Unlike SN 1993J, whose companion star was detected \citep{mau04}, the direct observation of the precursor
of SN 2008ax in pre-explosion HST images does not (at present) allow us to definitely discriminate between a single star or a binary system.
In the single-star scenario, the progenitor had a relatively high-mass (8-9M$_\odot$) C/O core
and a final mass (C/O core + He/H envelope) in the range 11-13M$_\odot$ \citep{cro08}. 
The mass of the C/O core was a factor of two times higher than that of the star that generated SN 1993J. 
This would imply that the progenitor of SN 2008ax was a star with a main sequence mass of about 
25-30M$_\odot$, significantly higher than the 12-17 M$_\odot$ estimated for SN~1993J. 
However, such a high C/O mass estimated for SN 2008ax would be expected to affect the shape of its light curve, 
making it broader than what is observed. This raises a potential problem in the interpretation of the observed 
SN evolution.

Alternatively, the progenitor could have been an initially less massive star (10-14M$_\odot$) in a binary 
system, embedded in a coeval cluster. In this case
other objects are expected to contaminate the magnitude and colour estimates of the source at the position 
of the progenitor in the images analysed by \citet{cro08}. If this is true, multiple sources near the 
SN position will be eventually recovered in future observations of the explosion site.

\section{Conclusions}
SN 2008ax has a number of observational properties in common with other type IIb SNe. 
In particular the shape of the optical light curves, the bolometric luminosity and the spectral line velocities
resemble those of SN 1993J. This suggests for the two events similar explosion and ejecta parameters.
However, the evolution of the spectra of SN 2008ax toward a totally H-deprived (Ib) spectral type is faster than in
SN 1993J.
In addition, SN 2008ax does not show the prominent early-time optical peak related to the shock breakout, 
and has slightly bluer colours compared with SN 1993J. This indicates that the two SNe are similar, 
but not identical.
 
Two different progenitors could produce SNe like 2008ax: a single, high-mass 
star that lost mass via strong stellar winds to become a massive WNL or,
alternatively, a less-massive star that had its envelope stripped away through interaction with a companion star 
\citep{cro08}.
The latter scenario agrees better with the SN evolution. 

Detailed modelling of the SN data is required to provide a robust estimate of the explosion and ejecta parameters, 
which is key information to finally unveil the nature of the WR star that exploded as SN 2008ax.

\section*{Acknowledgments}
This manuscript is partly based on observations collected at the Hale Telescope, Palomar Observatory, as part of a collaborative agreement 
between the California Institute of Technology, its divisions Caltech Optical Observatories and the Jet Propulsion 
Laboratory (operated for NASA), and Cornell University.
The paper is also based on observations obtained at the 60-inch Telescope of the Palomar Observatory and 2-m Liverpool Telescope.
The Liverpool Telescope is operated on the island of La Palma by Liverpool John Moores University in the 
Spanish Observatorio del Roque de los Muchachos of the Instituto de Astrofisica de Canarias with financial 
support from the UK Science and Technology Facilities Council.
We thank John Dann of the Wise Observatory staff for his expert
assistance with the observations.\\
This work, conducted as part of the award "Understanding the lives of
massive stars from birth to supernovae" (S.J. Smartt) made under the
European Heads of Research Councils and European Science Foundation
EURYI (European Young Investigator) Awards scheme, was supported by
funds from the Participating Organisations of EURYI and the EC Sixth
Framework Programme. S.J.S also thanks the Leverhulme Trust for
funding through the Philip Leverhulme Prize scheme.\\
The work of D.S. was carried out at Jet Propulsion Laboratory,
California Institute of Technology, under a contract with NASA. A.G. acknowledges the Benoziyo Center for 
Astrophysics and the William Z. and Eda Bess Novick New Scientists Fund at the Weizmann Institute
of Science.\\
We acknowledge the usage of the HyperLeda database (http://leda.univ-lyon1.fr).

\end{document}